\documentclass[12pt]{article}
\usepackage{epsfig}
\usepackage{amsfonts}
 \newcommand{\beq}{\begin{equation}}
 \newcommand{\eeq}{\end{equation}}
 \newcommand{\beqa}{\begin{eqnarray}}
 \newcommand{\eeqa}{\end{eqnarray}}
 
\begin{document}

\begin{center}

%
%
{\Large{\bf The direct link model for polymer rings as a topological field theory and the second topological
    moment in dense systems  }}\\[2cm]

{\large{ Matthias Otto}}\\
 Institut f\"ur Theoretische Physik, 

Universit\"at G\"ottingen, Bunsenstrasse 9, D-37073 G\"ottingen,
Germany \footnote{Present address: SPEC - CEA, Orme des Merisiers, F-91191 Gif sur Yvette, France. } \\ [2cm]

\end{center}

\underbar{Abstract}
Polymer rings in solution are either permanently entangled or not. 
Permanent topological restrictions give rise to additional entropic
interactions apart from the ones
arising due to mere chain flexibility or excluded volume. Conversely, entangled polymer
rings systems may be formed by closing randomly entangled flexible
linear chains. The dependance of linking numbers between randomly entangled
rings on the chain length, more specifically the second topological
moment $\langle n^2 \rangle$, i.e. the average squared linking number, may
be determined. In this paper, an approach recently discussed in mathematical physics and called abelian BF theory,
is presented which allows to express the linking constraint in its
simplest form, the Gauss integral, in terms of two gauge fields. The
model of Brereton and Shah for a single ring entangled with many other
surrounding rings is rederived. The latter model is finally used to
calculate the second topological moment, in agreement with a recent
result by Ferrari, Kleinert, and Lazzizzera obtained by using ${\rm n}$-component $\phi^4$ theory in the
limit ${\rm n}\rightarrow 0$. 


\newpage
\section{Introduction}
The topology of knots and links has attracted the interest of
mathematicians and physicists alike (see for a general introduction
e.g.\cite{kauffman:93,guada:93,gilbert:94,kleinert:95,
nechaev:96,kholo:98}). In polymer physics, the role of
topological constraints has been discussed since its beginnings \cite{wasserman:60,frisch:62}. On
the level of a self-avoiding walk description for the polymer conformation,  the
Gauss invariant (defined below) was discussed by Edwards who also considered the role
of higher order link invariants (HOLC). In \cite{edwards:68}, the abelian
Chern-Simons theory (however not under the present term which is taken from the authors of \cite{chern:74}) was used to express the self-linking number of a
single ring. In his discussion of HOLC, Edwards pointed to the role of a 3-vertex connecting 3 gauge
fields, a remark anticipating the non-abelian
Chern-Simons theory which was used by Witten \cite{witten:89} some 20 years later to obtain polynomial
invariants for knots and links. 
These algebraic invariants which are usually defined by mathematical
knot theorists in terms of recursion
relations (skein relations) with respect to the crossings appearing in
the 2-dimensional projections of knots and links have
also been applied to polymers (for this important approach which will
not explored here, the reader is referred to \cite{kholo:98}).
The conjecture by Edwards has been validated by the work of 
Cotta-Ramusino, 
Guadagnini, et al. (see  \cite{guada:93,kholo:98} and references therein) and by Axelrod et al., and by Alvarez and Labastida (see \cite{alvarez:93} and references therein), who show that the Gauss invariant and all HOLC appear in a
perturbation series of the non-abelian Chern-Simons theory when the
vacuum expectation value of products of Wilson loops is evaluated.
This result suggests to treat the Gauss invariant as an approximate
measure of the true topology.

The concept of the abelian variant of 
Chern-Simons field theory was reconsidered by many authors, among them
\cite{kholo:94,brereton:95,otto:diss,otto:98,ferrari:99,ferrari:00}, 
also for many rings systems, as
it arises naturally whenever the Gaussian linking number is enforced
as a constraint in the partition function. Partial reviews of the
subject are \cite{nechaev:96,kholo:98,kleinert:95}. The non-abelian variant has
been discussed only recently by some authors \cite{kholo:94,kleinert:95,otto:diss}, however its
implementation in terms of a well motivated physical model for
entangled polymers is not so clear. 

Whenever the abelian Chern-Simons theory is used in the
context of the Gauss integral, an important problem arises in terms of
self-linking integrals, where the double line integration is not taken
with respect to 
two different ring contours but with respect to a single ring contour. 
These integrals attain a meaning only when using the so-called
vertical framing, i.e. by replacing the single ring by a band. For
details and the relevant literature we refer the reader to \cite{guada:93}. 

Following a suggestion of Thompson (see e.g. the review  \cite{blau:91}),
 an alternative topological quantum field theory, termed BF theory \cite{bf}, is
 used in the present paper. 
In contrast to Chern-Simons theory, it avoids the
appearance of self-linking integrals, and was
introduced, at least to the knowledge of the author, to the subject of polymer topology in \cite{otto:diss}. Recently,
BF theory was rediscovered in 
\cite{ferrari:00,ferrari:00b} which is also concerned about avoiding self-linking integrals. In agreement with existing literature on topological quantum field theory as done in \cite{blau:91}, the term BF theory will be used to underline the difference to Chern-Simons theory explained above.

In the present paper, the direct link model for a system of $n_p$
rings of \cite{brereton:80} is used. It singles out a single ring, say $C_1$ and enforces a
linking number $n$ of this test ring with all other rings $C_\beta\in \{C_2,\dots,C_{n_p}\}$ in the
system. The Gauss invariant expressing the linking number as a double
line integral is then represented in terms of an abelian
BF theory. Next, the conformational degrees of freedom of all rings but
$C_1$ are summed over, giving rise to a ``mass term'' to be added to
the BF theory. Then, one gauge field is integrated over, and the field
theory of Brereton and Shah \cite{brereton:80} is obtained, expressed in terms of one
remaining gauge field and the conformational field ${\bf r}^1
(s)$, where $s$ parametrizes the polymer contour. This theory is then
used to evaluate what has recently been called the second topological
moment \cite{ferrari:00}, $\langle n^2 \rangle$.
Essentially, the purpose of this paper is two-fold: one, to inform
polymer physicists of a new approach (BF theory) to tackle entangled
polymers, and two, to present a very simple derivation of the second
topological moment in dense systems.

The outline of the paper is as follows. In section 2, the direct
link model is presented and expressed as an abelian BF theory. The
model of Brereton and Shah is derived. In section 3, the second
topological moment is evaluated. Finally, in the concluding section
the relation to similar work is discussed and a simple scaling
argument is evoked to interpret the result.

\section{The direct link model as a BF theory}
Let the conformation of the single test ring and all other rings 
be given by their coordinates $\{{\bf r}^1 (s)\}$ and $\{{\bf r}^\beta (s)\}$ where 
$\beta=2,\dots, n_p$ respectively.
Now let $n$ 
be the linking number of the ring 
 $C_1$ with all other rings
 $C_\beta$
\beq
n=\sum_{\beta\neq 1}^{n_p}\Phi(C_1, C_\beta).
\eeq
where 
\beq
\Phi(C_\alpha, C_\beta)=\frac{1}{4\pi}\oint_{C_\alpha}\oint_{C_\beta}
d{\bf r}^\alpha\wedge d{\bf r}^\beta\cdot
\frac{{\bf r}^\alpha-{\bf r}^\beta}{|{\bf r}^\alpha-{\bf r}^\beta|^3}
\eeq
is the Gaussian linking number. As discussed in the literature
\cite{kauffman:93,guada:93}, the double line integral on the r.h.s. is a topological
invariant, i.e. is invariant with respect to all deformations of the rings
given by their coordinates $\{{\bf r}^\alpha\}$ and $\{{\bf
  r}^\beta\}$ which do not alter the topological state. However, this
invariant is not unique. A counterexample is the Whitehead link \cite{kauffman:93} which represents a
linked state between two rings and has linking number $\Phi(C_\alpha,
C_\beta)=0$, the same value which results for a pair of unlinked
rings. However, the Whitehead is maintained by self-interactions of
one ring with itself. For random walk rings, these self-interactions
effectively disappear, and thus the Whitehead link is unknotted to a 
pair of unlinked rings.

The partition function of the system with a given linking number
$n$ reads as:
\beq
\label{b1}
Z(n)=Z_0\left\langle
\delta(n,\sum_{\beta\neq 1}^{n_p}\Phi(C_1, C_\beta))
\right\rangle_{\{{\bf r}^1 (s)\},\{{\bf r}^\beta (s)\}}
\eeq
The  Kronecker delta appearing in Eq.(\ref{b1}) 
may be expressed in terms of the integral
\beq
\delta(n,f)=\int_{-\pi}^\pi \frac{dg}{2\pi}e^{ign-igf}
\eeq
giving rise to a topological ``charge'' $g$ conjugate to $n$.
One obtains the corresponding conjugate partition function $Z(g)$:
\beq
\label{b2}
Z(g)=\left\langle
\exp\left(
ig\sum_{\beta\neq 1}^{n_p}\Phi(C_1, C_\beta)
\right)
\right\rangle_{\{{\bf r}^1 (s)\},\{{\bf r}^\beta (s)\}}
\eeq
The parameter $g$ may therefore be considered as chemical potential
for linking numbers.
In what follows, the conformational model for the rings is restricted
to closed random walk chains. The excluded volume effect is
neglected. As discussed in \cite{otto:diss}, this simplification
restores the uniqueness of the Gauss invariant for links.

The argument inside the average in 
 Eq.(\ref{b2}) 
may now be expressed (in the spirit of a Hubbard-Stratonovich
transformation) in terms of an abelian BF theory as follows \cite{blau:91}:
\beqa
\label{b3}
Z(g)&=&{\cal N}\int {\cal D}{\bf A}\int {\cal D}{\bf B}
\exp\left(i\int {\bf B}\wedge d{\bf A}\right)\nonumber\\
&&\left\langle \exp\left(ig\oint_{C_1}d{\bf r}^1 \cdot{\bf A}\right)
\right\rangle_{\{{\bf r}^1 (s)\}}
\left\langle \exp\left(i\int {\bf B}\cdot{\bf j}\right)
\right\rangle_{\{{\bf r}^\beta (s)\}}
\eeqa
The constant ${\cal N}$ is a normalization. The wedge product $\wedge$
is an exterior product which generates the algebra of differential
forms over $\mathbb{R}^3$, $\Omega^*(\mathbb{R}^3)$ \cite{nakahara:90}. In this language ${\bf A}$
and ${\bf B}$ are
1-forms, and the invariant expression ${\bf B}\wedge d{\bf A}$ is a
3-form. In terms of local coordinates it reads as
$\epsilon_{\lambda\mu\nu}B_\lambda\partial_\mu A_\nu dx^1\wedge
dx^2\wedge dx^3$. Therefore, no volume element appears in the
integration. Line integrations are specified explicitly.
The notation given above is widely used in the field
theory literature \cite{blau:91}, and is used here to make manifest
the coordinate invariance of the field theory. Later on, when
evaluating 2-point functions, we return to explicit coordinate
representations of the fields.
The variable  ${\bf j}({\bf x})$ is the tangent vector density, simply called the
``current'', of the chains
$\beta=2..n_p$
\beq
{\bf j}({\bf x})=\sum_\beta\oint_0^N ds \dot{\bf r}^\beta (s) 
\delta({\bf x}- {\bf r}^\beta (s)).
\eeq

The conformational averages with respect to the test ring $C_1$
and with respect to the other rings factorize for random walk chains. The conformation of the
surrounding rings are represented in terms of the current ${\bf
  j}$. In fact, they have fused to a single effective ring. 
As a reminder, the appearance of self-linking numbers has been avoided.

Next, the conformational coordinates for the surrounding rings
$C_1$ for $\beta\neq 1$ are summed over. Then, the functional
integral with respect to the gauge field ${\bf B}$ is performed.
First, the average with respect to the ring chains 
 $C_\beta$ 
is carried out. For random walk rings one obtains:
\beq
\label{b4}
\left\langle \exp\left(i\int {\bf B}\cdot{\bf j}\right)
\right\rangle_{\{{\bf r}^\beta (s)\}}
=\exp\left(-\frac{1}{2}\int d{\bf x}\int d{\bf x}'{\bf B}_\mu ({\bf x}){\bf B}_\nu ({\bf x}')
\langle {\bf j}_\mu({\bf x}){\bf j}_\nu ({\bf x}')\rangle\right).
\eeq
On the r.h.s. of this equation the chain index has been omitted. 
The average of the correlation function for the currents at points
${\bf x}$ and ${\bf x}'$ is carried out in the limit of very long
rings. 
Let us denote the number of segments per ring by $N$ and the average
segment length by $l$.
Neglecting $1/N$ terms arising due to the closure constraint \cite{brereton:92} for the surrounding rings, the
current-current correlation function is given by the following
expression (the same approximation is made in \cite{brereton:80}):
%
%
\beq
\label{b5}
\langle {\bf j}_\mu({\bf x}){\bf j}_\nu ({\bf x}')\rangle=
\delta_{\mu\nu}\frac{\rho l^2}{d}\delta({\bf x}-{\bf x}')
\eeq
where
$\rho=(n_p N)/V$ is the average segment density. Eq.(\ref{b5}) is an approximation valid for sufficiently concentrated systems. The question may be raised for these systems whether the Gauss invariant is still valid. Certainly the probability for more complicated links, such as the Borromean rings (a 3 ring link which falls apart when cutting one ring), which are only detected by HOLC, is higher but remains low compared to the Gauss invariant measuring the pairwise entanglement of rings. In order to determine the second topological moment given in the next section, these considerations are irrelevant as the Gauss invariant is simply an observable of the system (see below).

Let 
$G=\rho l^2/3$, 
then the partition function reads as
\beqa
\label{b6}
Z(g)&=&{\cal N}\int {\cal D}{\bf A}\int {\cal D}{\bf B}
\exp\left(i\int {\bf B}\wedge d{\bf A}
-\frac{G}{2}\int {\bf B}\cdot{\bf B}
\right)\nonumber\\
&&\left\langle \exp\left(ig\oint_{C_1}d{\bf r}^1 \cdot{\bf A}\right)
\right\rangle_{\{{\bf r}^1 (s)\}},
\eeqa
where ${\cal N}$ is an adjusted normalization factor.
Now the integration with respect to the gauge fields
 ${\bf B}$ 
may be carried out, which yields the effective model of
 Brereton und Shah \cite{brereton:80}:
\beq
\label{b7}
Z(g)={\cal N}\int {\cal D}{\bf A}
\exp\left(
-\frac{1}{2G}\int (\nabla\wedge{\bf A})^2
\right)
\left\langle \exp\left(ig\oint_{C_1}d{\bf r}^1\cdot {\bf A} \right)
\right\rangle_{\{{\bf r}^1 (s)\}}
\eeq
Leaving aside for the moment the average with respect to the ring 
 $C_1$ and using the language of quantum field theory,
one is left with a so-called Wilson loop that is averaged with respect to
a euclidean Yang-Mills theory in $2+1$ dimensions. Wilson loops are
considered in quantum field theory to study the confinement
problem. An analogy to this problem has been used in \cite{otto:96}
to study the collapse transition of randomly entangled polymer
rings in 2D (for earlier studies see \cite{nechaev:93,rosta:93}).

\section{The second topological moment}
The second topological moment $\langle n^2 \rangle$ for a single test ring entangled with
many other rings given the restriction that all conformational
averages are taken with respect to random walks may now be determined rather
simply.
Following \cite{ferrari:00,ferrari:00b} $\langle n^2 \rangle$ may be determined from:
\beq
\langle n^2 \rangle=-\frac{\partial^2}{\partial g^2}Z(g)|_{g=0}
\eeq
More generally $Z(g)$ is the generating function for all topological moments.
It is formulated for the present purpose as follows:
\beq
\label{b8}
Z(g)=\left\langle \exp\left(ig\oint_{C_\alpha}d{\bf r}^\alpha {\bf A} \right)
\right\rangle_{\{{\bf A}\},\{{\bf r}^\alpha (s)\}}
\eeq
The normalization factor has been absorbed into the average with
respect to ${\bf A}$. To separate the gauge fields and the
conformational coordinates, the line integral in Eq.(\ref{b8}) is
expressed as follows:
\beq
\label{b8a}
\oint_{C}d{\bf r}\cdot {\bf A}=\int_{\bf k}
A_\mu({\bf k})\oint_{C}ds\;\dot{
  r}_\mu(s)e^{i{\bf k}\cdot{\bf r}(s)}
\eeq
The summation convention is understood. The abbreviation $\int_{\bf
  k}$ represents $\int d^3k/(2\pi)^3$.
In order to perform the functional integral with respect to ${\bf A}$,
a gauge has to be  used, which in the present case is $\nabla\cdot{\bf
  A}=0$.
In this case the 2-point correlation function for ${\bf A}$ in Fourier
space is given by:
\beq
\label{b9}
\langle A_\mu({\bf k})A_\nu({\bf q})\rangle_{\{{\bf A}\}}
=\delta({\bf k}+{\bf q})\frac{G}{k^2}\left(\delta_{\mu\nu}-\hat{k}_\mu\hat{k}_\nu\right)
\eeq
where $\hat{k}_\mu=k_\mu/|{\bf k}|$ are the components of a unit vector.
In order to perform the conformational average, the only expression
needed is
\beq
\Phi_{\mu\nu}(s_1,s_2;{\bf k})=
\langle \dot{r}_\mu (s_1)\dot{r}_\mu (s_2)e^{i{\bf k}\cdot({\bf r}(s_1)-{\bf r}(s_2))}\rangle
\eeq
This average has been evaluated for random walks rings
\cite{brereton:92} and reads as follows:
\beqa
\label{b10}
\Phi_{\mu\nu}(s_1,s_2;{\bf k})&=&
\left[
\frac{l^2}{3}\delta_{\mu\nu}\left(\delta(s_1-s_2)-\frac{1}{N}\right)+\frac{l^4}{9N}k_\mu
k_\nu |s_1-s_2|\left(1-\frac{|s_1-s_2|}{N}\right)\right]
\nonumber\\
&\times&\exp\left[-\frac{l^2k^2}{6}|s_1-s_2|\left(1-\frac{|s_1-s_2|}{N}\right)\right]
\eeqa
The $k$-dependent terms in the first factor on the RHS of this
equation do not give any contributions due to gauge invariance. This is most obvious in the Landau
gauge, ${\bf k}\cdot{\bf A}({\bf k})=0$.

Now, the second topological moment is given by (using a summation
convention for repeated indices):
\beq
\langle n^2 \rangle=-\left\langle
\int_{\bf k}\int_{\bf q}
A_\mu({\bf k})A_\nu({\bf q})\oint ds_1 \oint ds_2
\dot{
  r}_\mu(s_1)\dot{
  r}_\nu(s_2) e^{i{\bf k}\cdot{\bf r}(s_1)}e^{i{\bf q}\cdot{\bf r}(s_2)}
\right\rangle_{\{{\bf A}\},\{{\bf r} (s)\}}
\eeq
Using the correlators (\ref{b9}) and (\ref{b10}), one obtains the
expression:
\beqa
\label{b11}
\langle n^2 \rangle&=&G\int_{\bf k}\frac{1}{k^2}\oint ds_1 \oint ds_2
\left[
\frac{l^2}{3}(\delta_{\mu\mu}-1)\left(\delta(s_1-s_2)-\frac{1}{N}\right)\right]
\nonumber\\
&\times&\exp\left[-\frac{l^2k^2}{6}|s_1-s_2|\left(1-\frac{|s_1-s_2|}{N}\right)\right]
\eeqa
The last equation gives rise to 2 terms. The first stems from the
delta function $\delta(s_1-s_2)$ and is readily evaluated. The
second needs further analysis: if $s_1=s_2$, the exponential
equals $1$, and one needs to evaluate an integral in ${\bf k}$ space
which depends on a cutoff $\Lambda\sim \xi^{-1}$; if $s_1\neq s_2$,
the integration with respect to $s_1$, $s_2$ gives a non-trivial
contribution to the integration in ${\bf k}$-space, the details of
which are discussed in the appendix. The result is when inserting for $G$
\beq
\label{b12}
\langle n^2 \rangle=\rho l^3\left(
\frac{Nl\Lambda}{9\pi^2}-\frac{\sqrt{N}}{\sqrt{2(3\pi)^3}}-\frac{l\Lambda}{9\pi^2}
\right)
\eeq
where $\Lambda=\xi^{-1}$ is an inverse cutoff length.
As the result has been obtained in the limit $N\gg 1$, the last
term which is independant of $N$ may be dropped, so the final result in
the large $N$ limit reads as:
\beq
\label{b13}
\langle n^2 \rangle\simeq\rho l^3\left(
\frac{Nl\xi^{-1}}{6\pi^2}-\frac{\sqrt{N}}{\sqrt{3(2\pi)^3}}
\right)
\eeq
Apart from the numerical prefactors, the result is identical to the
one derived from an ${\rm n}$-component field theory in the limit
${\rm n}\rightarrow 0$ in \cite{ferrari:00}.
The leading scaling term in Eq.(\ref{b13}) was also found by Brereton and Shah \cite{brereton:82} who calculated the generating function $Z(g)$ directly in the conformational space without the use of gauge fields, using a pre-averaging procedure as a further approximation. Their result contains however the square root of the ratio $l/\xi$ in contrast to the present result which is a minor difference and might be due to the approximation used in their work.

\section{Discussion}
The topological constraint of fixed linking number of a test ring
entangled with $n_p$ surrounding rings has been implemented within the
framework of a random walk model for the polymer conformation, by
using the simplest link invariant, the Gauss integral. The latter one
has been reformulated in terms of a topological quantum field theory,
the so-called abelian BF theory. Two results have been obtained: first,
the model of Brereton and Shah \cite{brereton:80} has been derived, second, the second
topological moment in the limit of large segment numbers $N\gg 1$ has
been calculated, in agreement with the previous result by Ferrari, Kleinert, and Lazzizzera \cite{ferrari:00,ferrari:00b}. 
The same result may also be obtained from
directly averaging the squared Gauss integral with respect to a random
walk conformation, a work which will be presented elsewhere. The gauge
field approach, however, is conceptually very appealing as it
separates topological interactions from conformational entropy.

The present method avoids the complicated ${\rm n}$-component field
theory used in \cite{ferrari:00}. The latter approach is, however, better suited
to treat the full problem including excluded volume interactions.
On the other hand, the Gauss invariant is ambiguous for self-avoiding
walk rings. 
Considering work by Moroz and Kamien \cite{moroz:97} it remains
an open problem whether a topological field coupled to the ${\rm
  n}$-component field for the polymer conformation changes the random
walk result significantly. 
The authors consider self-avoiding walks
with writhe. In fact, a chemical potential for writhe is introduced
which gives the coupling constant for the interaction between a
topologocial gauge field and conformational ${\rm
  n}$-component field, in a very 
 similar
way as  in
\cite{ferrari:00}. The scaling behavior of the radius of gyration and
the first two moments of the writhe are calculated. As to the first
question, the effect of writhe is found to be irrelevant (to
one-loop)
, i.e. the self-avoiding walk fix point corresponding to the exponent
$\nu\simeq 0.588$ remains unchanged.
Essentially this result is due to the fact that a topological field
theory lacks a scale.
Concerning moments of the writhe (which is similar to calculating
moments of $n$ as done above), no scaling dependence on the number
of polymer segments is found which is a surprising result. A clearification of this issue is
certainly necessary.
Let us note that Kholodenko and Vilgis \cite{kholo:96} calculated the writhe of semiflexible polymers which is found to scale as $N^{1/2}$.

Finally let us interpret the result Eq.(\ref{b13}) in terms of a simple
picture. Obvious the second topological moment basically scales as
$\langle n^2 \rangle\sim \rho l^3 N$ as $N$ becomes large. A simple argument, also
discussed in \cite{vilgis:97}, gives a similar result. Let the average segment
density $\rho$ inside the test ring be given by the number of rings
crossing the interior of the test ring, which is equal to the average
linking number per ring $\bar{n}$, times the number of segments $N$, divided by
the volume of the test ring $R^3$. Then the density reads as 
\beq
\rho=\frac{\bar{n}N}{R^3}
\eeq
Assuming that the segment density in the interior of the test ring is
equal to the segment density elsewhere in the system, and employing
$R^3\sim l^3 N^{3/2}$, i.e. using the Gaussian result for the ring volume,
one immediately obtains
\beq
\bar{n}=\rho l^3\sqrt{N}
\eeq
Both $\bar{n}$ and $\sqrt{\langle n^2 \rangle}$ scale as $\sqrt{N}$,
so the theoretical calculation presented above reproduces the
characteristic scaling with chain length derived from the simple argument.
The difference between density-dependent prefactors appears to be
superficial. In fact
 in the case of dense
melts where $\rho\sim l^{-3}$ and which is the regime where the above
argument applies, the prefactors coincide, and $\bar{n}=\sqrt{\langle n^2 \rangle}$.\\
\\
{\bf Acknowledgements}:
Discussions with F. Ferrari and H. Kleinert are gratefully acknowledged. 
The author is grateful to T.A. Vilgis who introduced him to the subject of polymer topology.
The author thanks M. Weigt for a critical
reading of the manuscript, and acknowledges financial support by the DFG under
grant Zi209/6-1.

\section*{Appendix}
In the following, the calculation of the r.h.s. of Eq.(\ref{b11})
\beqa
\label{b11.app}
\langle n^2 \rangle&=&\frac{2}{3}G\int_{\bf k}\frac{1}{k^2}\oint ds_1 \oint ds_2
\left[
l^2\left[\delta(s_1-s_2)-\frac{1}{N}\right)\right]
\nonumber\\
&\times&\exp\left[-\frac{l^2k^2}{6}|s_1-s_2|\left(1-\frac{|s_1-s_2|}{N}\right)\right]
\eeqa
is given in more detail. The delta function in the integrand gives
rise to the integral
\beqa
I_1&=&\frac{2}{3}GNl^2\int_{\bf
  k}\frac{1}{k^2}=\frac{2}{3}GNl^2\frac{4\pi}{(2\pi)^3}\int_0^\Lambda
dk\nonumber\\
&=&GNl^2\frac{\Lambda}{3\pi^2}
\eeqa
where the ${\bf k}$ integration has been carried out up to the cutoff
parameter $\Lambda\sim \xi^{-1}$.
The $1/N$ term in the integrand on the r.h.s. of Eq.(\ref{b11.app})
is given by
\beq
I_2=-\frac{2}{3}G\frac{l^2}{N}\int_{\bf k}\frac{1}{k^2}\oint ds_1 \oint ds_2
\exp\left[-\frac{l^2k^2}{6}|s_1-s_2|\left(1-\frac{|s_1-s_2|}{N}\right)\right]
\eeq
Concerning the integration with respect to $s_1$ and $s_2$ in the last
equation, two cases need to be considered: $s_1=s_2$
and $s_1\neq s_2$. In the first case, the exponential function gives
$1$ and the integration gives a factor of $N$, the result being
\beq
I_{2a}=-Gl^2\frac{\Lambda}{3\pi^2}
\eeq
The case $s_1\neq s_2$ is slightly more involved. First, let us
integrate with respect to $k$, which gives:
\beq
I_{2b}=-G\frac{l^2}{N}\frac{4\pi}{(2\pi)^3}\oint ds_1 \oint ds_2
\sqrt{2\pi}\left[\frac{l^2}{3}|s_1-s_2|\left(1-\frac{|s_1-s_2|}{N}\right)\right]^{-1/2}
\eeq
Now the domain of integration with respect to $s_1$ and $s_2$ may be restricted
to two times one half the original one giving:
\beq
I_{2b}=-\frac{2}{3}G\frac{l^2}{N}\frac{4\pi}{(2\pi)^3}2\int_0^N ds_1 \int_0^{s_1} ds_2
\sqrt{2\pi}\left[\frac{l^2}{3}(s_1-s_2)\left(1-\frac{(s_1-s_2)}{N}\right)\right]^{-1/2}
\eeq
A shift of variables $u=(s_1-s_2)/N$, $v=(s_1+s_2)/N$ is easily performed and
leads to 
\beq
I_{2b}=-\frac{2}{3}G\frac{l^2}{N}\frac{4\pi}{(2\pi)^3}\sqrt{2\pi}
N^2\int_0^1
du \int_a^{2-a} dv
\left[\frac{l^2}{3}Nu\left(1-u\right)\right]^{-1/2}
\eeq
From now on the integration is elementary, and one obtains
\beq
I_{2b}=-GlN^{1/2}\frac{1}{\sqrt{6\pi^3}}
\eeq
Adding $I_1$, $I_{2a}$, and $I_{2b}$ and inserting for $G=\rho l^3/d$
for $d=3$ gives the r.h.s. of Eq.(\ref{b11}) in the main text.


\newpage
\begin{thebibliography}{999}
\bibitem{kauffman:93}
L.H. Kauffman,
{\it Knots and physics}, World Scientific, Singapore 1993.

\bibitem{guada:93}
E. Guadagnini,
{\it The link invariants of the Chern-Simons field theory}, Walter de
Gruyter, Berlin 1993.

\bibitem{gilbert:94}N.D. Gilbert, T. Porter, {\it Knots and surfaces},
  Oxford University Press, Oxford 1994.

\bibitem{kleinert:95}
H. Kleinert, {\it Path integrals in quantum mechanics, statistics, and
  polymer physics}, 2nd ed., World Scientific, Singapore 1995.  

\bibitem{nechaev:96}
S.K. Nechaev,
{\it Statistics of knots and entangled random walks}, World
Scientific, Singapore 1996.

\bibitem{kholo:98}
A.L. Kholodenko, T.A. Vilgis,
Phys. Rep. {\bf 298}, 251 (1998).

\bibitem{wasserman:60}
E. Wasserman,
J. Am. Chem. Soc. {\bf 82}, 4433 (1960).

\bibitem{frisch:62}
H.L. Frisch, E. Wasserman,
J. Am. Chem. Soc. {\bf 83}, 3789 (1962).

\bibitem{edwards:68}
S.F. Edwards, J. Phys. A {\bf 1}, 15 (1968).

\bibitem{chern:74}
S.S. Chern, J. Simons, Ann. Math. {\bf 99}, 48 (1974).

\bibitem{witten:89}
E. Witten, Commun. Math. Phys. {\bf 121}, 351 (1989).

\bibitem{kholo:94}
A.L. Kholodenko, T.A. Vilgis,
J. Phys. (Paris) {\bf 4}, 843 (1994).

\bibitem{alvarez:93}
M. Alvarez, J.M.F. Labastida, Nucl. Phys. {\bf 395}, 198 (1993); {\it ibid.} {bf 433}, 555 (1995).

\bibitem{brereton:95}
M.G. Brereton, T.A. Vilgis, J. Phys. A: Math. Gen. {\bf 28}, 1149 (1995).

\bibitem{otto:diss}
M. Otto, ``Die statistische Mechanik flexibler, verschlaufter und
unverschlaufter Polymerringe'', Ph.D. thesis, University of Mainz,
1996. Available at {\it http://www.Theorie.Physik.UNI-Goettingen.DE/$\sim$otto/pubs.html}.

\bibitem{otto:98}
M. Otto, T.A. Vilgis, Phys.Rev.Lett {\bf 80}, 881 (1998).

\bibitem{ferrari:99}
F. Ferrari, I. Lazzizzera,
Nucl. Phys. {\bf B539}, 673 (1999);
Phys. Lett. {\bf B444}, 167 (1998).

\bibitem{ferrari:00}F. Ferrari, H. Kleinert, I. Lazzizzera,
Phys. Lett. {\bf A 276}, 31 (2000), cond-mat/0002049;
{\it Calculation of second topological moment $\langle m^2\rangle$ of
  two entangled polymers}. Preprint, cond-mat/0003355, available at
{\it http://xxx.lanl.gov}.  

\bibitem{ferrari:00b}
F. Ferrari, H. Kleinert, I. Lazzizzera,
{\it Field theory of $N$ entangled polymers}. Preprint,
cond-mat/0005300, available at
{\it http://xxx.lanl.gov}. 


\bibitem{blau:91}
M. Blau, G. Thompson,
Ann. Phys. (N.Y.), {\bf 205}, 130 (1991).

\bibitem{bf} The term ``BF'' originates from its non-abelian form
  whose general action is given by (wedge) product of a field ${\bf
    B}$ and field tensor ${\bf F_A}$ (corresponding to another field
  ${\bf A}$), $S=\int Tr({\bf B}\wedge {\bf F_A})$. In the abelian
  case, ${\bf F_A}=d{\bf A}$.

\bibitem{brereton:80}
M.G. Brereton, S. Shah,
J. Phys. A: Math. Gen. {\bf 13}, 2751 (1980).

\bibitem{nakahara:90}
M. Nakahara, {\it Geometry, topology and physics}, Adam Hilger,
Bristol 1990.

\bibitem{otto:96}
M. Otto, T.A. Vilgis, 
J. Phys. A: Math. Gen. {\bf 29}, 3893 (1996).

\bibitem{nechaev:93}
S.K. Nechaev, V.G. Rostiashvili,
J. Phys. II France {\bf 3}, 91
(1993).

\bibitem{rosta:93}
V.G. Rostiashvili, S.K. Nechaev, T.A. Vilgis,
Phys. Rev. E {\bf 48}, 3314 (1993).

\bibitem{brereton:92}
M.G. Brereton, T.A. Vilgis, Phys. Rev. A {\bf 45}, 7413 (1992).

\bibitem{brereton:82}
M.G. Brereton, S. Shah, J. Phys. A: Math. Gen. {bf 15}, 985 (1982).

\bibitem{vilgis:97}
T.A. Vilgis, M.Otto, Phys.Rev. E Rap.Comm. {\bf 56} R1314 (1997).



\bibitem{moroz:97}
J.D. Moroz, K.D. Kamien, 
Nucl.Phys. B{\bf 506} 695 (1997).

\bibitem{kholo:96} 
A.L. Kholodenko, T.A. Vilgis, J. Phys. A: Math. Gen.{\bf 29}, 939 (1996).

\end{thebibliography}
\end{document}